\documentclass{PoS}

\title{Unbalanced low-density neutron matter}

\ShortTitle{Unbalanced low-density neutron matter}

\author{\speaker{Alexandros Gezerlis}\\
ExtreMe Matter Institute EMMI,
GSI Helmholtzzentrum f\"ur Schwerionenforschung GmbH, 64291 Darmstadt, Germany\\
Institut f\"ur Kernphysik,
Technische Universit\"at Darmstadt, 64289 Darmstadt, Germany\\
        E-mail: \email{gezerlis@theorie.ikp.physik.tu-darmstadt.de}}
        
\author{Rishi Sharma\\
        TRIUMF, 4004 Wesbrook Mall, Vancouver, BC, V6T 2A3, Canada,\\
        E-mail: \email{rishi@triumf.ca}}

\abstract{We consider polarized neutron matter at low densities. We have
performed Diffusion Monte Carlo simulations for normal neutron matter with different
population numbers for each species. We analyze the competition between different
phases in the grand canonical ensemble and mention aspects of neutron-star phenomenology
that are impacted by the effects described.}

\FullConference{Xth Quark Confinement and the Hadron Spectrum,\\
		October 8-12, 2012\\
		TUM Campus Garching, Munich, Germany}

\begin{document}

\section{Introduction}
Systems which feature $s$-wave pairing between fermions of different species
--- say $\uparrow$ and $\downarrow$ fermions --- exhibit interesting phase
structure as we vary the difference in their chemical potentials,
$\delta\mu=|(\mu_\uparrow-\mu_\downarrow)/2|$. For small enough $\delta\mu$,
the free energy gain provided by Cooper pairing drives the system into a
superfluid state. All fermions are paired up, and the number densities of
$\uparrow$ particles ($n_\uparrow$) and $\downarrow$ particles ($n_\downarrow$)
are equal. As we increase $\delta\mu$, pairing is stressed since in the
absence of pairing the corresponding Fermi surfaces would be split and the
system would be polarized, $\delta n = n_\uparrow-n_\downarrow\neq0$. At large enough
$\delta\mu$ the pairing is disrupted and the system exists in a normal phase.
~\footnote{Here we are assuming that the masses of the two species are equal. A
similar stress arises in systems of fermions with different masses (for
eg.~\cite{Gezerlis:2009}).} 

Stressed pairing of this kind shows up in a variety of physical systems. In
quark matter at densities relevant to the phenomenology of neutron stars, the
strange quark mass gives rise to a splitting between the effective quark
chemical potentials~\cite{Alford:1999}. In cold atoms, polarization can be
created by trapping different numbers of fermions of the two hyperfine states
that pair. Since the number of particles in the traps is fixed, polarization
leads to phase separation~\cite{Bedaque:2003} which can be
observed~\cite{Shin:2006} by looking at their density profile. 

In these proceedings, we will focus on the pairing between spin $\uparrow$ and
$\downarrow$ neutrons in the inner crust of neutron stars.  Dipole interactions
with the magnetic field split the chemical potentials of the two. At a
critical value of the field, pairing breaks, and there is a transition to the
normal phase from the superfluid phase. 

Calculating the critical field is relevant for the phenomenology of neutron-star 
crusts because the transport properties of the two phases are very
different. The superfluid phase can transport heat efficiently via Goldstone
bosons (although mixing between the superfluid Goldstone bosons and the lattice
phonons suppresses this effect~\cite{Cirigliano:2011,Chamel:2012}).
Furthermore, the normal phase features gapless neutrons near the Fermi surface,
and consequently has a large specific heat.  Both these effects tend to
increase the time scales~\cite{Brown:2009} associated with heat transport in
the inner crust if a significant fraction of the crustal neutrons are in the 
normal phase. 

The phase diagram as a function of $\delta\mu$ is well understood in BCS
theory. The normal phase is favored over the superfluid phase for
$\delta\mu=\delta\mu_c>\Delta_0/\sqrt{2}$ where $\Delta_0$ is the pairing gap
for $\delta\mu=0$. In a small window near $\delta\mu_c$ exotic phases called
LOFF~\cite{LOFF:196465} phases are theoretically favored.

Similar results can be argued for ``weak coupling''~\cite{Sharma:2008rc}: a regime where the
separation between the particles is much greater than the scattering length.
More formally, defining $k_F = (3\pi^2(n_\uparrow+n_\downarrow))^{1/3}$, and
the scattering length as $a$, the free energy of the normal and the superfluid
phases can be written in terms of functions that become smaller as $|k_Fa|$
decreases, if $k_Fa$ is small.  

For strongly coupled Fermi systems, however, we don't know of a controlled
expansion technique to calculate thermodynamic properties. Therefore, one needs
to perform intensive numerical calculations to compute the energies of both
the normal and the superfluid phase.

In these proceedings we will present QMC calculations of the energy of the
competing phases for three number densities
$n_1=6.65\times10^{-4}$ fm$^{-3}$, $n_2=2.16\times10^{-3}$ fm$^{-3}$, and
$n_3=5.32\times10^{-3}$ fm$^{-3}$ for different polarizations. For
comparison, these correspond respectively to $0.0042n_{\rm{sat}}$,
$0.0135n_{\rm{sat}}$, and $0.033n_{\rm{sat}}$ where
$n_{\rm{sat}}=0.16$ fm$^{-3}$ is the saturation density. The neutron-neutron
interaction has a large scattering length ($a\simeq-18.6$ fm) compared to 
the effective range $r_e$, $\sim2.7$ fm. We see that at the densities we consider $k_Fa$ is
large in magnitude ($k_Fa=5.02$, $7.44$, and $10.05$ corresponding to the 
three densities which are all deep in the strongly coupled regime). For
reference, we note that $k_Fr_e=0.59$, $0.88$, and $1.19$.

Phase competition is analyzed very simply in the grand canonical ensemble. If
we know the pressure of the competing phases as a function of the average
chemical potential $\mu=(\mu_\uparrow+\mu_\downarrow)/2$ and the splitting
$\delta\mu$, then the phase with the larger pressure at given $\mu$ wins.  At
the phase transition the pressure and $\mu$ should match. ($\delta\mu$ need
not be the same between coexisting phases, as long as $\delta\mu<\Delta$, where
$\Delta$ is the superfluid gap.)

The direct evaluation of the pressure as a function of $\mu$ is very difficult
because of the sign problem. For a fixed $n_\uparrow$, $n_\downarrow$ a
creative technique known as fixed node QMC~\cite{Carlson:Morales:2003} has been
developed to calculate the energy of fermionic systems ${\cal{E}}(n, \delta
n)$ accurately. We will describe this calculation next.

\section{Calculation of ${\cal{E}}$~\label{section:QMC}}

\begin{figure}[t]
\vspace{0.5cm}
\begin{center}
\includegraphics[width=0.46\textwidth]{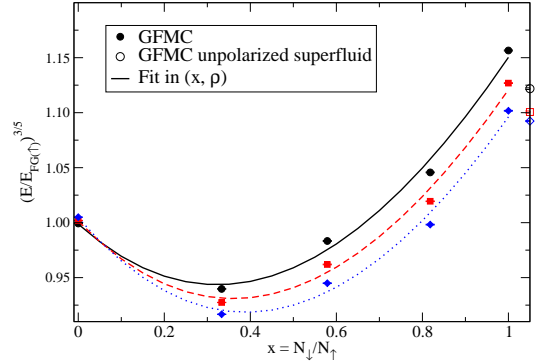}
\caption{Ground-state energy per particle (in units of the free Fermi gas energy)
scaled to the power of $3/5$ for normal spin-polarized neutron matter. 
Shown are QMC results at three different total densities $n_1, n_2, n_3$ 
(as points: black circles, red squares, blue diamonds, respectively), along
with fits to the Monte Carlo results using the set of functions of $(x, \rho)$
given in Eq.~$3.6$.  (as lines: black solid, red dashed, blue dotted,
respectively).  Also shown using hollow symbols are the results for an
unpolarized superfluid.}
\label{fig:ener}
\end{center}
\end{figure}

The heart of the calculation is the accurate evaluation of the energy as a
function of density and polarizations~\cite{Gezerlis:2011gq,Gezerlis:2011}. 
We use the Diffusion Monte Carlo (DMC) method which is expected to be exact, modulo
the fermion-sign problem. Schematically, this entails taking a variational
wave function $\Psi_{V}$ that encodes knowledge about the physics of the system
(e.g. is it superfluid, does it contain trimers, etc.). We first perform
a Variational Monte Carlo simulation and then project out the exact, 
lowest-energy eigenstate $\Psi_{0}$ from the trial (variational) wave 
function $\Psi_{V}$. In other words, we treat the Schr\"{o}dinger equation 
as a diffusion equation in imaginary time and evolve the variational 
wave function up to the limit of $\tau \rightarrow \infty$. 
This can be seen by expanding $\Psi_{V}$ in terms of the complete 
set $\Psi_{i}$ of the exact eigenstates of the Hamiltonian (with eigenvalues $E_i$):
\begin{eqnarray}
\Psi(\tau) & = &
 e^{-(H-E_T) \tau} \Psi_V
  =  \sum_i \alpha_i e^{-(E_i - E_T) \tau} \Psi_i \nonumber \\
 &=& \alpha_0 e^{-(E_0-E_T) \tau} \Psi_0, \hspace{1em} \lim \tau \rightarrow \infty~.
\label{evolution}
\end{eqnarray}
where $E_T$ is called the trial energy. In practice, the DMC technique is 
implemented by discretizing $\tau$. 

As already implied, the trial wave function is critically important, in 
that it is used as the fixed-node constraint. In addition, it is commonly also
used as an importance-sampling function. In this problem, we are interested
in simulating non-superfluid systems with unequal populations of spin-up
and spin-down neutrons. Thus, modulo a conceptually irrelevant (but numerically
useful) Jastrow term, the wave function describes the particles as being
in a free Fermi gas (two, to be precise, one for each spin-orientation). 
This formalism localizes all the correlations within the Jastrow
functions, leaving the rest of the wave function as a product of two
Slater determinants: 
\begin{equation} 
\Phi_S({\bf{R}}) =
{{\cal A}}[\phi_n(r_1) \phi_n(r_2) \ldots \phi_n(r_{N_{\uparrow}})] 
{{\cal A}}[\phi_n(r_{1'}) \phi_n(r_{2'}) \ldots \phi_n(r_{N_{\downarrow}'})]
\end{equation} 
where the unprimed (primed) indices correspond to
spin-up (spin-down) neutrons, $N_{\uparrow} + N_{\downarrow}' = N$, 
${\cal A}$ is
the antisymmetrizer, and the $\phi_n(r_k)$ are plane waves with periodic
boundary conditions inside a box of volume $L^3$. Using real numbers is often preferable numerically, so
it is customary to use linear combinations of the eigenfunctions. 
Shown below is the result for 7 particles:
\begin{equation}
\left |
\begin{array}{cccccc}
1 & 1 & \ldots &
1 \\
cos(2\pi x_1/L) & cos(2\pi x_2/L) & \ldots &
cos(2\pi x_7/L) \\
cos(2\pi y_1/L) & cos(2\pi y_2/L) & \ldots &
cos(2\pi y_7/L) \\
cos(2\pi z_1/L) & cos(2\pi z_2/L) & \ldots &
cos(2\pi z_7/L) \\
sin(2\pi x_1/L) & sin(2\pi x_2/L) & \ldots &
sin(2\pi x_7/L) \\
sin(2\pi y_1/L) & sin(2\pi y_2/L) & \ldots &
sin(2\pi y_7/L) \\
sin(2\pi z_1/L) & sin(2\pi z_2/L) & \ldots &
sin(2\pi z_7/L) \\\end{array}
\right |~~.
\end{equation}
The number 7 was chosen because it corresponds to a closed shell. Closed shells
exhibit no ambiguity in the selection of the momenta different particles are
placed in. 

In our simulations of polarized normal neutron matter we 
used 33 + 0, 57 + 19, 57 + 33, 33 + 27, and 33 + 33 particles. These 
correspond to relative fractions of $x =$ 0, 0.333, 0.579, 0.818, and 1,
respectively. At each
particle number combination, we performed simulations for three distinct total
number densities [$n =
(N_{\uparrow} + N_{\downarrow})/L^3$], as mentioned earlier. The
results of these computations are given in Fig.~\ref{fig:ener}, which 
shows the values of
\begin{equation}
g(x, n)= \left(\frac{{\cal{E}}(x, n)}{{\cal{E}}_{FG(\uparrow)}}\right)^{3/5}
~\label{eq:gN1}\;,
\end{equation}
where ${\cal{E}}_{FG(\uparrow)}=\frac{3}{5}\frac{(6\pi^2)^{2/3}}{2m}(n_{\uparrow})^{5/3}$
is the energy of a non-interacting gas of a single species.

\section{The free energy functions~\label{section:freeenergy}}
\subsection{The unpolarized superfluid phase}
The energy and the gap for the unpolarized superfluid phase for $n_1$, $n_2$
and $n_3$ were calculated in~\cite{Gezerlis:2008,Gezerlis:2010}. A convenient
parameterization for the data can be obtained by considering the dimensionless
function $g_{SF}(n)$, defined by the relation
\begin{equation}
{\cal{E}}(n)=\frac{3}{10}\frac{(3\pi^2)^{2/3}}{2m}(ng_{SF}(n))^{5/3}~\label{eq:gSF}\;,
\end{equation}
where, ${\cal{E}}$ is the energy density. The prefactors and the power of $n$
in the definition is chosen so that for a system of non-interacting
fermions, $g=1$. Since $g_{SF}$ is dimensionless, the density dependence in 
$g_{SF}$ can only arise in appropriate combinations with $a$ and $r_e$. In
particular, it motivates considering the form
\begin{equation}
g_{SF}(n) = g_{0, 0} + g_{0, 1}r_e n^{1/3}  + g_{1, 0} \frac{1}{n^{1/3} a} +
\cdot\cdot\cdot
\end{equation}
For the relevant densities, a good fit is obtained for $g_{0, 0}=0.942$, $g_{0,
1}r_e=0.003$MeV$^{-1}$, $g_{1, 0}/a=2.301$MeV. For comparison, for the
unitary Fermi gas ($a\rightarrow\infty$, $r_e\rightarrow0$), $g_{SF}$ is
independent of the density and is given by $(2\xi)^{3/5}$, where $\xi$ is the
Bertsch parameter. For $\xi=0.38$~\cite{Forbes:2011} calculated using similar
techniques, $(2\xi)^{3/5}=0.848$, which is close to $g_{0, 0}$. It is not
surprising that $g_{0, 0}$ differs from the unitary value since this
calculation is being done for neutron matter, and the interaction is more
complicated. One expects deviations also because the largest $k_Fr_e$ is
of the order of $1$.

With this parameterization, it is easy to calculate the pressure as a function
of the chemical potential using standard thermodynamical relations. 

\subsection{The normal phase}
The analysis of the normal phase is a little more involved because the energy
depends both on the density $n$ and the relative fraction
$x=n_\downarrow/n_\uparrow$ where we take $n_\uparrow$ to be the majority
species. A convenient generalization of Eq.~\ref{eq:gSF} is 
\begin{equation}
{\cal{E}}(x, n)=\frac{3}{5}\frac{(6\pi^2)^{2/3}}{2m}(n_{\uparrow}g(x,
n))^{5/3}~\label{eq:gN2}\;,
\end{equation}
which is equivalent to Eq.~\ref{eq:gN1}. The chemical potential is:
\begin{equation}
\mu = \frac{\partial {\cal{E}} }{\partial n} \Big |_{\delta n}\\
 = \frac{3^{2/3} \pi ^{4/3} \left(\frac{n }{x+1}\right)^{2/3}
   g(x,n)^{2/3} \left(\frac{2 n}{x+1}  \frac{\partial g}{\partial n}(x,n)+(1-x)
   \frac{\partial g}{\partial x}(x,n)+g(x,n)\right)}{2^{4/3} m}\;.
   ~\label{eq:mu}
\end{equation}

The chemical potential splitting is:
\begin{equation}
\delta\mu = \frac{\partial {\cal{E}} }{\partial \delta n} \Big |_{n}\\
=\frac{3^{2/3} \pi ^{4/3} \left(\frac{n }{x+1}\right)^{2/3}
   g(x,n)^{2/3} ((-x-1) \frac{\partial g}{\partial x}(x,n)+g(x,n))}
   {2^{4/3} m}\;.
   ~\label{eq:dmu}
\end{equation}

We fit the values of $g$ given in Fig.~\ref{fig:ener} with a linear 
combination of the following functions,
\begin{equation}
\{(x+1),\frac{(x+1)}{\rho^{1/3}},{(x+1)}{\rho^{1/3}},
(x+3)^2,\frac{(x+3)^2}{\rho^{1/3}},{(x+3)^2}{\rho^{1/3}},
(x+5)^3,\frac{(x+5)^3}{\rho^{1/3}},{(x+5)^3}{\rho^{1/3}}\}
~\label{eq:cubicinterp}
\end{equation}
The $x$ dependence is chosen so that at $\delta\mu=0$ for $x=1$.

The pressure is simply $P = -{\cal{E}}(n, \delta n) + \mu n +
\delta\mu \delta n$.

\section{Phase competition}

\begin{figure}[t]
\vspace{0.5cm}
\begin{center}
\includegraphics[width=0.46\textwidth]{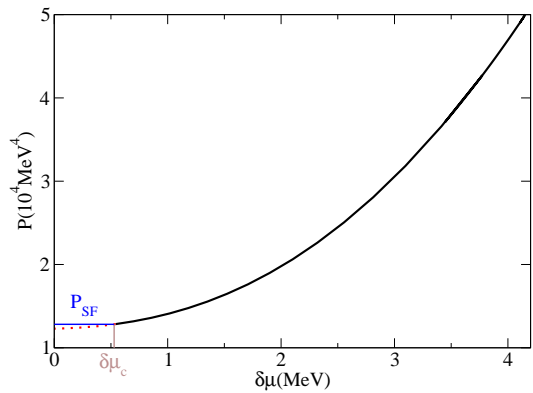}
\caption{Pressure as a function of $\delta\mu$ for fixed
$\mu=1.94$ MeV. For $\delta\mu>\delta\mu_c=0.53$ MeV the unpaired phase (bold
black curve) has a higher pressure. At $\delta\mu_c$ the $SF$ and the $NP$
phases can coexist and for smaller $\delta\mu$, the $SF$ phase is favored. On
the other extreme for $\delta\mu=4.18$ MeV, $x=0.0005$. We choose $\mu$ so that the
density of the normal phase at $\delta\mu_c$ is $n_2$.}
\label{fig:GrandCan}
\end{center}
\end{figure}

For a given $\mu$, $\delta\mu_c$ is simply the point where the pressure curves
of the unpolarized superfluid and the normal phases cross. For 
example, from the plot of the pressure as a function of 
$\delta\mu$ for $\mu=1.94$ MeV in Fig.~\ref{fig:GrandCan} we conclude that
$\delta\mu_c=0.53$ MeV. 

\begin{figure}[t]
\vspace{0.5cm}
\begin{center}
\includegraphics[width=0.46\textwidth]{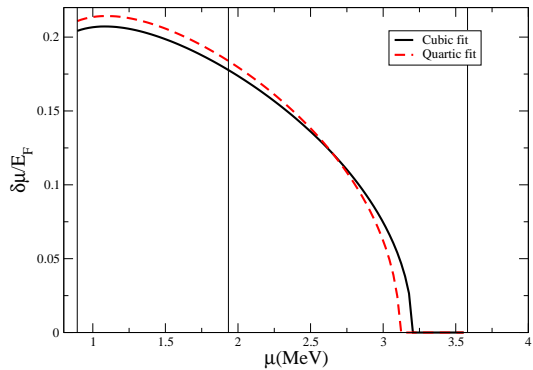}
\caption{Plot of $\delta\mu/E_F$ as a function of $\mu$. The two
curves correspond to two choices of interpolating functions in $x$. The solid
line (black) corresponds to the set of functions given in
Eq.~$3.6$, up to cubic power in $x$. The dashed line (red) 
corresponds to a larger set including quartic powers. The vertical
lines at increasing $\mu$ correspond to the chemical potentials at which the
densities of the unpolarized superfluid phase are $n_1$, $n_2$ and $n_3$
respectively.} \label{fig:dmuvsmu}
\end{center}
\end{figure}

In Fig.~\ref{fig:dmuvsmu} we show a plot of $\delta\mu_c/E_F$ as a function of
$\mu$, where $E_F=k_F^2/(2m)$ ($m$ is the mass of the neutron and  $k_F$ is the
Fermi momentum corresponding to the density in the superfluid phase). The range of
$\mu$ is chosen such that the density of the superfluid varies from $n_1$ at
the smallest value to $n_3$ at the largest. For reference we note that
$\Delta/E_F$ increases gradually from $0.4$ at $n_1$ to $0.44$ at
$n_3$~\cite{Gezerlis:2008}. This can be seen as a competition between
stronger pairing as $|k_Fa|$ increases weakened by increasing $k_Fr_e$. The
effect of the effective range on $\delta\mu_c/E_F$ is much more pronounced.  We
see that, not surprisingly, this makes it ``easier'' (in a dimensionless sense)
to break Cooper pairing at higher density. The two curves give an estimate of
the uncertainty associated with different interpolating functions of the same
data.  

\section{Conclusions}
To summarize, we have performed Diffusion Monte Carlo simulations for
polarized neutron matter at low density. Low density implies that the
interaction between neutrons is greatly simplified: three-body forces and
spin-orbit \& tensor interactions have a minimal effect, leading to a mainly
s-wave interaction. We have included the s-wave channel of the high-quality
phenomenological Argonne v18 potential, along with perturbative p-wave
corrections. The computations we have performed were undertaken for normal
neutrons of varying population numbers and are expected to provide tight upper
bounds to the true ground-state energies. 

Using these results we have extracted the critical chemical
potential splitting $\delta\mu_c$ at which there is a transition from the 
superfluid state to the normal state as we increase $\delta\mu$.

For a system with unequal number densities of $\uparrow$ and $\downarrow$
neutrons our calculations imply that the energetically favored state is that of
phase separation into an unpolarized superfluid and a polarized normal neutron
gas. This situation directly parallels what is found in systems of ultracold
fermionic atoms \cite{Pilati:2008} which have no effective-range and p-wave
contributions. The cold-atom system has also been probed experimentally and
quantitatively well-described by QMC calculations,
lending credence to the claim that our results for neutron matter are also
quite accurate. Using our simulations and worked published earlier, we have
extracted the critical relative fractions above which phase separation occurs
(for details see~\cite{Gezerlis:2011gq}). For relative fractions above the
critical values the competing phase that has been considered~\cite{Gezerlis:2011} is a 
homogeneous polarized superfluid state. The mixed phase is energetically
favored compared to the polarized superfluid state, meaning that we have
provided the lowest-energy configuration to date for the system under study. At
relative fractions lower than the critical value, our analysis suggests that the
system is entirely normal.

The quenching of pairing, if it materializes in a neutron star sufficiently
magnetized to polarize the neutrons, would have directly observable
consequences, as superfluidity impacts the specific heat and more generally
the thermal behavior of the star. This would also hold even for slightly
smaller magnetic fields. More generally, our predictions are in principle also
relevant to experiments with cold atoms which could in the future be performed
for finite effective ranges in the laboratory. \cite{Marcelis:2008} Such
experiments today make use of unequal spin populations as a matter of course.
Importantly, our microscopic simulations can also impact the terrestrial
phenomenology of nuclei. Skyrme-family energy-density functional theories of
nuclei have traditionally been fitted to experimentally measured masses of
nuclei, while also sometimes including infinite matter constraints. The latter
in the past consisted of the behavior of nuclear matter, while more recently
the neutron-matter pairing gap has also been used as an extra constraint.
\cite{Chamel:2008,Dutra:2012} Following this path to its logical consequences, a
recently completed work \cite{Forbes:2013} directly compares the neutron
polaron binding energy (i.e. the binding of one spin-down impurity embedded in
a sea of spin-up neutrons) with the predictions following from a group of
established Skyrme parametrizations. Extending this line of thinking, one
could generalize such comparisons to finite relative fractions and thus
provide a further constraint to nuclear density functionals: this all follows
from the fact that our DMC results provide a dependable microscopic solution
of the Schroedinger equation. Neutron-rich nuclei, in particular, are expected
to be impacted by the physics of polarized neutron matter.

There are many ways in which one could build on the calculations discussed in
this contribution. One could add an external periodic potential, which would
stand for the periodic lattice of neutron-rich nuclei found in a neutron-star
crust and could also teach us something about surface effects in neutron
drops. The physics of the static response of polarized (or, for that matter,
unpolarized) neutron matter is wholly uncharted from a microscopic
perspective. Any predictions on this system could also be extended to the
relevant setting of optical-lattice experiments, which could indirectly teach
us something about the neutron case, even without an effective range.
Similarly, one could extend calculations such as these to more species or even
finite-temperature effects, in an effort to directly probe thermal effects in
neutron-star crusts and cores.

\end{document}